\documentclass[letterpaper]{JHEP3}
\usepackage{epsfig}
\usepackage{bbm}
\def\beq{\begin{equation}}
\def\eeq{\end{equation}}
\def\bea{\begin{eqnarray}}
\def\eea{\end{eqnarray}}
\def\ba{\begin{array}}
\def\ea{\end{array}}

\def\part{\partial}

\title{The abelian and non-abelian Josephson effect and pseudo-goldstone bosons}

\author{ L.-P. Guay\thanks{E-mail: {\tt l-p-guay@umontreal.ca}}, \ R. B. MacKenzie \thanks{E-mail: {\tt rbmack@lps.umontreal.ca}}, \ M. B. Paranjape \thanks{E-mail: {\tt paranj@lps.umontreal.ca}} \\ 
Groupe de physique des particules, \\
Universit\'e de Montr\'eal \\
C.P. 6128, succ. centre-ville, Montr\'eal, \\
Qu\'ebec, CANADA H3C 3J7 }
\author{Paul Esposito \thanks{E-mail: {\tt esposito@physics.uc.edu}}, L. C. R. Wijewardhana 
\thanks{E-mail: {\tt rohana@physics.uc.edu}}\\ 
Department of Physics\\ 
University of Cinicinatti\\ 
P.O. Box 210011  Cincinatti, OH\\
 USA  45221-0011}

\abstract{ We formulate the Josephson effect in a field theoretic language which affords   a straightforward generalization to the non-abelian case.  We give some examples and apply the formalism to the case of $SO(5)$ superconductivity.
}


\preprint{UdeM-GPP-TH-05-141}

\dedicated{} 

\begin{document}

\newpage
\section{Abelian Josephson Effect}

The abelian Josephson effect \cite{Josephson} concerns  two macroscopic superconductors that are brought into weak contact with one another.  Each superconductor is described by a separate macroscopic state.   The two superconductors are brought into contact, and interact weakly with one another.  Physically, the wave function of the Cooper pairs, which have Bose  condensed to form the superconducting liquids  on either side of the junction, start to interact with one another.  The interaction allows for tunneling across the junction which is at the heart of the Josephson effect.  The effect can be succinctly described in terms of effective fields.   The formulation given by Feynman\cite{Feynman} is the most convenient.  

The superconductor on each side is described by one, macroscopic state $\psi (t)$ and $\chi (t)$ respectively.  The spatial dependence of the states, although important to define the volume occupied by the superconductor and indeed how it is positioned facilitating the interaction with the other side, is actually irrelevant for the effect.    With no coupling, the states each obey a free Schrodinger equation:
\begin{eqnarray}
i\hbar\partial_t\psi (t)=E_L\psi(t) \\
i\hbar\partial_t\chi (t)=E_R\chi(t) 
\end{eqnarray}
where $E_L$ and $E_R$ are the chemical potentials on either side.  These equations admit a doubled symmetry, $U(1)\times U(1)$, corresponding to independent rotations of the fields $\psi (t)\rightarrow e^{i\zeta}\psi (t)$ and $\chi (t)\rightarrow e^{i\eta}\chi(t)$. This enhanced symmetry is fictive, a consequence of treating the two superconductors as independent uncoupled entities.   Only the simultaneous identical rotation of both fields, the diagonal $U_D(1)$,  is truly a symmetry.  The phases $\theta$ and $\zeta$ may be taken to be time dependent, at the expense of adding the temporal component of a gauge field $A_0(t)$  with the appropriate transformation property.  In the superconducting state, each $U(1)$ symmetry is spontaneously broken giving rise to two Goldstone bosons\cite{Goldstone}.   The attendant Higgs mechanism gives the photon a mass, which physically manifests itself in the Meissner effect.  However, it should be stressed that there is actually only one photon; therefore only the diagonal $U_D(1)$ is gauged and the Goldstone boson associated with the spontaneous breaking of $U_D(1)$ is swallowed by the Higgs mechanism giving rise to the massive photon, while the Goldstone boson associated with the spontaneous breaking of the off-diagonal $U_{OD}(1)$ symmetry appears to be an undesirable consequence of treating the system as two independent superconductors.

Coupling the two superconductors together allows them to exchange charge with one another.  The coupled system is described by adding the simplest interaction which preserves the diagonal $U_D(1)$ symmetry.  Since the interaction is weak, the corresponding coupling constant is taken to be a small parameter.  If the coupling were not weak, it would be incorrect to treat the system as two coupled superconductors; we would just have one, albeit bigger, superconductor.  The coupling explicitly but softly, breaks the enhanced $U(1)\times U(1)$ to the diagonal $U_D(1)$.  This explicit symmetry breaking gives the putative second Goldstone boson a small mass and as such it is called a pseudo-Goldstone\cite{pseudo-Goldstone} boson.  The classic example of pseudo-Goldstone bosons in particle physics corresponds to the pions.  In QCD, the pions would be the Goldstone bosons arising from spontaneous breaking of chiral symmetry.  However, chiral symmetry is explicitly but softly broken by the quark mass terms.  Correspondingly,  the would-be massless Goldstone bosons acquire a small mass and are called pseudo-Goldstone bosons.  The simplest coupling of the two superconductors gives the equations:
\begin{eqnarray}
i\hbar\partial_t\psi (t)=E_L\psi(t)+K \chi(t)\\
i\hbar\partial_t\chi (t)=E_R\chi(t) +K\psi(t)
\end{eqnarray}
This system is  exactly solvable.  Writing $E_L=E+V$ and  $E_R=E-V$ the solution is
\beq
\left(\begin{array}{c}\psi(t)\\ \chi(t)\end{array}\right) =e^{-iEt/\hbar}\left(\cos\left(\omega t\right)-i\sin\left(\omega t\right)\left(\begin{array}{cc}{V\over\sqrt{V^2+K^2}}&{K\over\sqrt{V^2+K^2}}\\ {K\over\sqrt{V^2+K^2}}&{-V\over\sqrt{V^2+K^2}}\end{array}\right)\right)\left(\begin{array}{c}\psi_0\\ \chi_0\end{array}\right) 
\eeq
where $\omega = {\sqrt{V^2+K^2}\over\hbar}$.
The conserved charge in the system is $Q=\psi^*(t)\psi(t)+\chi^*(t)\chi(t)$, such that
$\dot Q=0$.  This means that the individual charges $Q_\psi=\psi^*(t)\psi(t)$ and $Q_\chi=\chi^*(t)\chi(t)$ satisfy $\dot Q_\psi =-\dot Q_\chi$. The overall charge of the junction is conserved, however due to tunnelling, the two superconductors can exchange charge. Indeed, calculating $Q_\psi$ yields
\bea\nonumber
&Q_\psi&=\psi^*(t)\psi(t)=\cos^2\left(\omega t\right)\psi^*_0\psi_0 \\ \nonumber +&\sin^2&\left(\omega t\right)({K\over\sqrt{V^2+K^2}}\chi^*_0+{V\over\sqrt{V^2+K^2}}\psi^*_0)({K\over\sqrt{V^2+K^2}}\chi_0+{V\over\sqrt{V^2+K^2}}\psi_0)\\ 
 &+&i\cos\left(\omega t\right)\sin\left(\omega t\right)\left(\left({K\over\sqrt{V^2+K^2}}\chi^*_0+{V\over\sqrt{V^2+K^2}}\psi^*_0\right)\psi_0 - c.c.\right).
\eea
Replacing $\psi_0=\sqrt\rho e^{i\theta_\psi}$ and $\chi_0=\sqrt\rho e^{i\theta_\chi}$ (the amplitude of the effective field is the same on both sides, the phase can differ) gives 
\beq
Q_\psi =\rho\left(\cos^2(\omega t) +\sin^2\left(\omega t\right)\left( 1+{2KV\over{V^2+K^2}} \cos(\theta_\psi-\theta_\chi)\right) -\sin (2\omega t){K\over\sqrt{V^2+K^2}}\sin(\theta_\psi-\theta_\chi)\right).
\eeq
There are two interesting cases to consider.  Firstly with $V=0$ we get the dc Josephson effect
\beq
Q_\psi =\rho\left(1 -\sin (2\omega t)\sin(\theta_\psi-\theta_\chi)\right).
\eeq
The Josephson current is given by
\beq
\dot Q_\psi =\rho\left(2\omega\cos (2\omega t)\sin(\theta_\chi-\theta_\psi)\right) .
\eeq
Writing $\omega =K/\hbar$ and noting that the dc current is valid for only short times, implying $\cos Kt/\hbar\approx 1$, yields
\beq
\dot Q_\psi =\rho\left(2K/\hbar \sin(\theta_\chi-\theta_\psi)\right)
\eeq
the familiar expression for the dc Josephson effect.

Secondly, for the ac effect we take $V>>K$ which yields
\beq
Q_\psi =\rho\left( 1 -{K\over V}\cos (2V t/\hbar+(\theta_\chi-\theta_\psi))\right)
\eeq
and consequently
\beq
\dot Q_\psi =\rho{2K\over\hbar}\sin (2V t/\hbar+(\theta_\chi-\theta_\psi)).
\eeq
The Josephson acceleration equation follows straightforwardly from the equations of motion for the time dependent phases $\psi(t)=\sqrt\rho_\psi (t) e^{i\theta_\psi(t)}$ and $\chi=\sqrt\rho_\chi(t) e^{i\theta_\chi(t)}$
\beq
\dot \theta_\chi(t)-\dot\theta_\psi(t) =2V/\hbar .
\eeq

An effective Lagrangian description of the situation is useful.  It is important to realize that it is not really a wavefunction, but a quantum field, that describes each superconductor; these fields are placed into interaction in a Josephson junction.  The effective Lagrangian in the above analysis is given by
\beq
{\cal L} = \psi^\dagger i\hbar\dot\psi +\chi^\dagger i\hbar\dot\chi -(\psi^\dagger \chi^\dagger )\left(\begin{array}{cc}{E+V}&0\\ 0&{E-V}\end{array}\right)\left(\begin{array}{c}\psi\\ \chi\end{array}\right) -(\psi^\dagger \chi^\dagger )\left(\begin{array}{cc}0&{K}\\ {K}&0\end{array}\right)\left(\begin{array}{c}\psi\\ \chi\end{array}\right) .
\eeq
In the absence of the coupling term, $K=0$, the symmetry of this model corresponds to independent phase transformations of the two fields $\psi\rightarrow e^{i\zeta}\psi$ and $\chi\rightarrow e^{i\eta}\psi$.  The fact that physically the amplitude on either side of the effective fields are equal and vary very little means that the $U(1)$ symmetry is spontaneously broken.  The photon, which we have not included in our analysis, will then absorb the attendant Goldstone boson and becomes massive giving the Meissner effect.  However, there really are not two independent photons.  The doubling of the symmetry is only an artefact of our effective description of two disjoint superconductors.  The field that describes the fluid of Cooper pairs is just one single albeit local field, that takes on one value over the position of one superconductor and another over the position of the second.  The true symmetry of the theory corresponds to $U(1)$ transformations of this field not the $U(1)\times U(1)$ symmetry that we have found.  The coupling of the two superconductors together explicitly breaks the symmetry $U(1)\times U(1)\rightarrow U(1)$.  Symmetries that spontaneously break give rise to massless Goldstone bosons.  Explicitly, but softly breaking the symmetry no longer produces Goldstone bosons, but slightly massive particles called pseudo-Goldstone bosons.  In the effective Lagrangian, the parameter $K$ is the soft breaking parameter.  The phase transformation that is equal and opposite on either side of the junction corresponds exactly to excitations in the direction of the pseudo-Goldstone bosons.  The frequency associated with these oscillations is correspondingly small, $\omega = {K\over \hbar}$.  The ac effect can be seen as an accumulation of the phase $(\theta_\chi-\theta_\psi)\rightarrow 2Vt/\hbar +(\theta_\chi-\theta_\psi)$.

\section{Non-abelian Josephson Effect}

The non-abelian Josephson effect can now be formulated, in terms of a junction of two effective systems which interact with one another very weakly.  Each system should have the same symmetry, or at least one symmetry should be a subset of the other.  Each system should exhibit spontaneous symmetry breaking.  For those generators that correspond to the same symmetry, this doubling of symmetry and the corresponding doubling of the Goldstone bosons that are produced, must be an artefact of the description. And indeed, coupling the two systems together, so that only the diagonal action of the symmetry generators is preserved, will give rise to pseudo-Goldstone bosons.  These excitations will correspond to the non-abelian generalization of the Josephson effect.

We are unaware of such a description of the Josephson effect either in condensed matter physics or in particle physics.  The only direct reference to pseudo-Goldstone bosons and the Josephson effect is in a paper of Zhang\cite{Zhang}, where he is considering the $SO(5)$ model for the high temperature superconductivity/anti-ferromagnet system.  However he does not consider junctions, and the pseudo-Goldstone bosons arising there are a consequence of explicit $SO(5)$ symmetry breaking terms that are added to the effective Lagrangian to push the system away from the $SO(5)$ invariant critical point.  The Josephson effect is considered in a paper by E. Demler et al\cite{Berlinsky} within the context of the $SO(5)$ theory; their analysis is complementary to ours.

There is some reference to non-abelian Josephson effect in the paper of Ambegoakar et al\cite{deGennes} which formulates the problem for the A phase of liquid Helium$^3$.  However the emphasis is not on symmetry considerations but on the geometrical and physical layout that could give rise to a junction.  In any case, we feel that in the condensed matter literature, there is some understanding that the Josephson effect, abelian or non-abelian, does correspond to the excitation of pseudo-Goldstone bosons, however it is not explicitly and simply spelled out, as we attempt to do here.  We will give two examples of the non-abelian generalization.  
\subsection{$SO(3)$ Model}
Here we present a model with $SO(3)$ symmetry as the common symmetry.  We take a complex doublet representation, $\psi$, on one side and a real triplet representation, $\vec\phi$, on the other.  The effective Lagrangian is given by
\beq
{\cal L}={\cal L}_\psi+{\cal L}_{\vec\phi}+{\cal L}_I
\eeq
with
\bea
{\cal L}_\psi&=&\dot\psi^\dagger\dot\psi -\lambda (\psi^\dagger\psi -a^2)^2\\
{\cal L}_{\vec\phi}&=&{1\over 2}\dot{\vec\phi}\cdot\dot{\vec\phi} -\gamma (\vec\phi\cdot\vec\phi -v^2)^2\\
{\cal L}_I&=&-K\psi^\dagger\vec\sigma\psi\cdot\vec\phi 
\eea
where $\vec\sigma$ are the usual Pauli matrices and $\lambda ,\gamma , a, K$ and $v$ are positive constants.  The symmetry of this model is given by $U(2)\times SO(3)$ when the interaction term is removed.   Then the symmetry will break spontaneously to $U(1)\times SO(2)$.  The complex doublet breaks the original $U(2)\sim U(1)\times SU(2)$, to a particular combination of one generator in the $SU(2)$ subgroup  and the generator of the $U(1)$ sub-group, not unlike the symmetry breaking in the standard model of the electro-weak interaction.  The real triplet spontaneously breaks the $SO(3)$ symmetry to $SO(2)$, the sub-group corresponding to rotations about the vacuum direction.

With the interaction term, the symmetry is reduced to $U(1)\times SO_D(3)$ where the subscript $D$ stands for the diagonal sub-group.  Some of the Goldstone bosons that are produced in the previous analysis of spontaneous symmetry breaking now become pseudo-Goldstone bosons.  To understand the breaking pattern, we must find the minimum of the potential (the part of the Lagrangian which does not depend on the time derivatives) and find the spectrum of the small oscillations about this minimum.  This is completely equivalent to finding the normal modes of a classical coupled system.  The Goldstone bosons will have zero frequencies, the pseudo-Goldstone bosons will have frequencies that vanish as the coupling constant of the interaction terms is taken to zero, and finally there will be massive modes, whose frequencies will be proportional to the other parameters of the model, and as such will correspond to arbitrarily heavy excitations in the limit that $\lambda$ and $\gamma$ go to infinity.  To find the location of the minimum, we take the variational derivative with respect to the fields and set it equal to zero:
\bea
2\lambda (\psi^\dagger\psi -a^2)\psi +K\vec\sigma\psi\cdot\vec\phi &=&0\\
4\gamma (\vec\phi\cdot\vec\phi -v^2)\vec\phi+K\psi^\dagger\vec\sigma\psi&=&0
\eea
Any solution of this system is mapped to another solution by the action of the group that is spontaneously broken.  Hence we can find one solution, and obtain all others related to it by an appropriate gauge transformation.  Since with a transformation within $SU(2)$ we can take $\psi$ from one point to any other point with the same amplitude, we can choose without loss of generality
\beq
\psi = \psi_R \left( \begin{array}{c}0\\1\end{array}\right)
\eeq
where $\psi_R$ is real.  Then the second equation implies that $\phi =\phi_3(0,0,1)$, where $\phi_3$ is of course real.  This yields two equations in two variables, which are tractable,
\bea
2\lambda (\psi_R^2 -a^2)\psi_R -K\psi_R\phi_3 &=&0\\
4\gamma (\phi_3^2 -v^2)\phi_3-K\psi_R^2&=&0.
\eea
These equations yield cubic equations for each variable.  These are easily solved using a computer, however, it is not very important to have the exact solution.  We are happy with a perturbative solution.  We take
\bea
\psi_R=a +\alpha K \\
\phi_3=v +\beta K.
\eea
Substituting in the equations and keeping terms to first order in $K$ yields
\bea
2\lambda (2a\alpha K) -Kv &=&0\\
4\gamma (2v\beta K)v-Ka^2&=&0.
\eea
We easily find $\alpha$ and $\beta$ to give
\bea
\psi_R=a + {v\over 4\lambda a}K \\
\phi_3=v +{a^2\over 8\gamma v^2} K.
\eea
Now we must find the second variation of the potential evaluated at this minimum, and then diagonalise it to find the eigenfrequencies and the normal modes.  It is easier to work with real fields, we write $\psi =\left(\begin{array}{c}\psi_{1R} +i\psi_{1I}\\ \psi_{2R} +i\psi_{2I}\end{array}\right)$. The minimum is found above at $\psi_{1R} =\psi_{1I}=\psi_{2I}=0$.  The matrix of second partial derivatives is $7\times 7$, however it is very sparse, and breaks up into 3 groups of $2\times 2$ matrices and one singlet.  We find to first order in $K$ (here $f_i$ represents any of the fields)
\beq
{\partial^2 V\over \partial f_i\partial f_j}=\left(\begin{array}{ccccccc}4Kv&0&0&0&2Ka&0&0\\ 0&4Kv&0&0&0&-2Ka&0\\ 0&0&8\lambda a^2 +4vK&0&0&0&-2Ka \\ 0&0&0&0&0&0&0 \\ 2Ka&0&0&0&Ka^2/v&0&0 \\ 0&-2Ka&0&0&0&Ka^2/v&0 \\ 0&0&-2Ka&0&0&0&8\gamma v^2+3Ka^2/v\end{array}\right).
\eeq
Diagonalization of the matrix reveals the following frequencies and corresponding eigenvectors (not normalized):
\bea
\omega &=& 0; \psi_{2I}\sim v_1\\
\omega &=& 0; (\psi_{1R},\phi_1)= (a, -2v)\sim v_2\\
\omega &=& (4v+a^2/v)K;(\psi_{1R},\phi_1)= (2v, a)\sim v_3\\
\omega &=& 0; (\psi_{1I},\phi_2)= (a, 2v)\sim v_4\\
\omega &=& (4v+a^2/v)K;(\psi_{1I},\phi_2)= (-2v, a)\sim v_5\\
\omega &=& 8\lambda a^2; (\psi_{2R},\phi_3)= (1, 0)\sim v_6\\
\omega &=& 8\gamma v^2;(\psi_{2R},\phi_3)= (0, 1)\sim v_7\\
\eea
The final two eigenvalues and eigenvectors $v_6$ and $v_7$ have corrections of $\circ (K)$ but these are simply uninteresting, these modes correspond to the massive radial oscillations, and they decouple in the limit $\lambda ,\gamma\rightarrow\infty$.  The interesting modes are the light modes, i.e. the massless Goldstone modes and the light pseudo-Goldstone modes.  We observe that there are three Goldstone modes and two pseudo-Goldstone modes.  
The uncoupled system is expected to have five Goldstone bosons, one for each spontaneously broken symmetry.  However, two of those modes no longer correspond to a symmetry once the interaction is added.  The two independent rotations of the triplet that do not leave $\phi_3$ invariant are no longer symmetries.  Only if we act simultaneously on the complex doublet  by the $SU(2)$ matrix corresponding to the same group element, we preserve a symmetry.  These correspond to the two Goldstone bosons $v_2$ and $v_4$. The transformations corresponding to $v_3$ and $v_5$ rotate the fields $\vec\phi$ and $\psi$ in opposite directions in group space and no longer correspond to symmetries.  Hence they give rise to  pseudo-Goldstone bosons.  The transformation corresponding to a rotation about $\phi_3$ of just the field $\vec\phi$ and of the transformation of just the field $\psi$ that is the combination of the $U(1)$ and the $SU(2)$ generator which leave $\psi_{2R}$ invariant are still  symmetries.   They are not spontaneously broken and do not give rise to Goldstone or pseudo-Goldstone bosons.

\subsection{$SO(5)$ Model}
In this section we will analyze the $SO(5)$ model that has been put forward by Zhang\cite{Zhang} to describe in a unified framework the anti-ferromagnetic and superconducting phases of the high temperature superconductors.  The system is described by a real five component order parameter ${\vec\varphi}$.  The first two components $\vec\phi\equiv (\varphi_1,\varphi_2)$ are the real and imaginary parts of a complex field and describe the superconductivity.  The last three components $\vec n\equiv (\varphi_1,\varphi_2,\varphi_3)$ correspond to a real triplet field which is the order parameter for the anti-ferromagnetism.  The five vector is usually normalized to unity, however, we will not impose this constraint from the beginning.  We will obtain it as we decouple the massive excitations by taking certain coupling constants to infinity.  The critical point is described by an $SO(5)$ invariant effective Lagrangian, and corresponds to the point in the phase diagram where we have the coexistence of the the superconductivity and the anti-ferromagnetism. It is described by a Lagrangian of the symmetry breaking type
\beq
{\cal L}_{SO(5)}(\vec\varphi)= {1\over 2}\dot{\vec\varphi}\cdot\dot{\vec\varphi} - \lambda ({\vec\varphi}\cdot{\vec\varphi}-a^2)^2
\eeq
Modifiying the doping of the material can drive the system into the superconducting phase or the anti-ferromagnetic phase.  The critical point corresponds exactly to a half-filled band, adjusting the doping pushes the system into one phase or the other.  At the level of the effective Lagrangian we add the explicit soft symmetry breaking terms
\beq
{\cal L_{\rm doping}}(g, {\vec\varphi})= -g(\vec\phi\cdot\vec\phi - \vec n\cdot\vec n).
\eeq
For $g$ positive, this potential drives the minimum into the vector $\vec n$ hence anti-ferromagnetic while for $g$ negative the minimum is in the vector $\vec\phi$ hence superconducting.  Zhang had imposed the constraint that the five vector is normalized and had added only the first term or the second term to drive the system into one phase or the other.  We find our treatment is completely equivalent, and due to the enhanced symmetric treatment of the $\vec\phi$ variables and the $\vec n$ variables, more easily tractable.  Irrespective of these considerations, the doping terms break the symmetry explicitly from $SO(5)\rightarrow SO(3)\times SO(2)$.  The $SO(3)$ symmetry describes the spin-rotation symmetry broken in the anti-ferromagnetic phase while the $SO(2)$ describes the phase rotational symmetry broken in the superconducting phase.  

Without the doping terms, the spontaneous symmetry breaking is from $SO(5)\rightarrow SO(4)$.  As $SO(5)$ has ten generators while $SO(4)$ has six, this would give rise to  four Goldstone bosons.  The excitations of the real five vector would correspond to four massless modes and one massive mode, obviously coming from oscillations of the orientation and of the length respectively of the order parameter.  With the explicit symmetry breaking terms, on the superconducting side, only one Goldstone boson while three pseudo-Goldstone bosons  are produced.  This is because the generator of $SO(2)$ is spontaneously broken and a symmetry of the theory giving rise to the one Goldstone boson, while the three generators of the $SO(3)$ symmetry are part of the unbroken $SO(4)$ symmetry of the $SO(5)$ symmetric situation and do not give rise to any Goldstone bosons. The three other generators that would be spontaneously broken in the $SO(5)$ symmetric situation, are explicitly broken here, and give rise to pseudo-Goldstone bosons.  On the anti-ferromagnetic side, the $SO(3)$ symmetry is spontaneously broken to $SO(2)$ giving rise to two Goldstone bosons.  The unbroken $SO(2)$ is part of the unbroken $SO(4)$ symmetry (when $SO(5)$ is not explicitly broken) as are the generators of the $SO(2)$ which acts on the superconductor variables $\vec\phi$  and do not give rise to any Goldstone or pseudo-Goldstone bosons.  Hence there are two generators of $SO(5)$ which would rotate $\vec n$ into $\vec\phi$ which are explicitly broken here and give rise to two pseudo-Goldstone bosons.  Zhang in his article\cite{Zhang} is referring to these pseudo-Goldstone bosons, which in fact have nothing to do with the Josephson effect.

A Josephson type junction is modeled with the consideration of two independent systems described by a $SO(3)\times SO(2)$ invariant Lagrangian  for each system.  This combined system would have a doubled symmetry $(SO(3)\times SO(2))\times (SO(3)\times SO(2))$ and spontaneous symmetry breaking would give a double number of Goldstone bosons and pseudo-Goldstone bosons, depending on the case as in the previous paragraph.  The addition of an interaction term that preserves only the diagonal $SO(3)\times SO(2)$ symmetry and would give rise to a number of new pseudo-Goldstone bosons, which would be responsible for the Josephson tunneling between the two systems. The Lagrangian of the system is
\beq
{\cal L}_{\rm Josephson)}={\cal L}_{SO(5)}({\vec\varphi_1})+{\cal L_{\rm doping}}(g_1,{\vec\varphi_1})+{\cal L}_{SO(5)}({\vec\varphi_2})+{\cal L_{\rm doping}}(g_2,{\vec\varphi_2})+ K \vec\varphi_1\cdot\vec\varphi_2
\eeq
The coupling term is the simplest term invariant under the diagonal $SO_D(5)$ symmetry.  The choice of a positive coupling constant $K$ drives the system to want to align the two five vectors $\vec\varphi_1$ and $\vec\varphi_2$.  If we pick the parameters $g_1,g_2$ on both sides to be superconducting, for example, then Josephson effect that we would describe, would be a straightforward generalization of the usual abelian Josephson effect between two superconductors.  Similarly, a junction between two anti-ferromagnets would yield a straightforward application of the analysis of the non-abelian Josephson effect of the previous sub-section.  Hence we would like to consider the junction corresponding to a superconductor on one side but an anti-ferromagnet on the other taking $g_1>0$ while $g_2<0$.  As the coupling term between the two systems respects the diagonal  $SO(3)\times SO(2)$ invariance, the position of the minimum of the potential can be moved arbitrarily via the action of this group.  This will have no effect on the spectrum of oscillations about the minimum.  It is the position of the amplitude of the vectors $\vec\phi_1,\vec n_1\vec\phi_2\vec n_2$ which are more interesting.  As we are interested in Josephson tunneling between the superconductor and the antiferromagnet we can reduce the model further to include the dynamics for just the norms of these four vectors.  Using the notation $|\vec\phi_1|=\phi_1$, $|\vec\phi_2|=\phi_2$, $|\vec n_1|=n_1$, $|\vec n_2|=n_2$ we get the following effective Lagrangian
\bea\nonumber
{\cal L}_{\rm Josephson,\,\, reduced}&=&{1\over 2}(\dot{\phi_1}^2+\dot{n_1}^2) - \lambda ({\phi_1}^2+n_1^2-a^2)^2 +{1\over 2}(\dot{\phi_2}^2+\dot{n_2}^2) - \lambda ({\phi_2}^2+n_2^2-a^2)^2\\ \nonumber &-&g_1(\phi_1^2-n_1^2)+g_2(\phi_2^2-n_2^2)\\ &+& K( \phi_1\phi_2 +n_1n_2).
\eea
We take, for convenience, the same parameters $\lambda$ and $a$ on both sides, this corresponds to an identical system on either side before the explicit symmetry breaking is added with the constants $g_1,g_2$.   All the parameters are positive.  The explicit symmetry breaking drives the system to the anti-ferromagnetic side for the variables 1 and to the superconducting side for the variable 2.  The equations for the minimum are
\bea
(4\lambda ({\phi_1}^2+n_1^2-a^2)+2g_1)\phi_1 -K\phi_2=0\\
(4\lambda ({\phi_2}^2+n_2^2-a^2)-2g_2)\phi_2 -K\phi_1=0\\
(4\lambda ({\phi_1}^2+n_1^2-a^2)-2g_1)n_1 -Kn_2=0\\
(4\lambda ({\phi_2}^2+n_2^2-a^2)+2g_2)n_2 -Kn_1=0.
\eea
These equations have the following matricial form 
\bea
\left(\ba{cc} A+2g_1&-K\\ -K&B-2g_2\ea\right)\left(\ba{c}\phi_1\\ \phi_2\ea\right)=0\\
\left(\ba{cc} A-2g_1&-K\\ -K&B+2g_2\ea\right)\left(\ba{c}n_1\\ n_2\ea\right)=0
\eea
where $A=4\lambda ({\phi_1}^2+n_1^2-a^2)$ and $B=4\lambda ({\phi_2}^2+n_2^2-a^2)$.  To have a non-trivial solution the two determinants must vanish:
\bea
(A+2g_1)(B-2g_2)-K^2=0\\
(A-2g_1)(B+2g_2)-K^2=0
\eea
Adding and subtracting these two equations gives
\bea
AB=K^2+4g_1g_2\\
Ag_2-Bg_1=0
\eea
which has the solution 
\beq
\left(\ba{c} A\\ B\ea\right)=\pm\sqrt{4+(K^2/g_1g_2)}\left(\ba{c} g_1\\ g_2\ea\right).\label{sq}
\eeq
Thus the values of $A$ and $B$ are fixed functions of the coupling constants $K,g_1,g_2$ and are independent of the coupling constant $\lambda$, especially in the limit which decouples the massive excitations, $\lambda\rightarrow\infty$.  The solution for the minimum then is
\bea
\phi_1={K\over A+2g_1}\phi_2\\
n_1={K\over A-2g_1}n_2
\eea
with the values of $\phi_2$ and $n_2$ determined from by self consistency
\beq
\ba{lcl}A=&4\lambda \left(\left({K\over A+2g_1}\phi_2\right)^2+\left({K\over A-2g_1}n_2\right)^2-a^2\right)&=\sqrt{4+(K^2/g_1g_2)}g_1\\
B=&4\lambda ({\phi_2}^2+n_2^2-a^2)&=\sqrt{4+(K^2/g_1g_2)}g_2\ea .
\eeq
Since the variables represent the amplitudes of the superconducting or anti-ferromagnetic order parameter, they must be positive.  This requires $A\pm2g_1>0$, which in turn requires $A>0$.  Hence the positive sign is chosen in equation \ref{sq}.
This system is an inhomogeneous system of linear equations for the variables $\phi_2^2$ and $n_2^2$ which can be written in matrix form as
\beq
\left(\ba{cc}({K\over A+2g_1})^2&({K\over A-2g_1})^2\\ 1&1\ea\right)\left(\ba{c}\phi_2^2\\ n_2^2\ea\right) =\left(\ba{c}{\sqrt{4+(K^2/g_1g_2)}g_1\over 4\lambda }+a^2\\ {\sqrt{4+(K^2/g_1g_2)}g_2\over 4\lambda }+a^2\ea\right) .
\eeq
The solution is obtained by multiplying by the inverse of the $2\times 2$ matrix, this yields
\beq
\left(\ba{c}\phi_2^2\\ n_2^2\ea\right) ={K^2g_1\over -8Ag_2^2}\left(\ba{cc}1&-({K\over A-2g_1})^2\\ -1&({K\over A+2g_1})^2\ea\right)\left(\ba{c}{\sqrt{4+(K^2/g_1g_2)}g_1\over 4\lambda }+a^2\\ {\sqrt{4+(K^2/g_1g_2)}g_2\over 4\lambda }+a^2\ea\right) .
\eeq
The solution for $\phi_2$ and $n_2$ requires taking the square root.  Again the positivity of these fields imply that the positive root must be taken.  The solutions are valid only for a range of the parameters, they are certainly valid for $K<<g_1,g_2$.  In the other limit, $K>>g_1,g_2$ they are only valid for $g_1\approx g_2$.  We are not interested in the solution for arbitrary values of the coupling constants but in the large $\lambda$ limit.  In this limit we get
\beq
\left(\ba{c}\phi_2^2\\ n_2^2\ea\right) ={K^2g_1\over -8Ag_2^2}\left(\ba{cc}1&-({K\over A-2g_1})^2\\ -1&({K\over A+2g_1})^2\ea\right)\left(\ba{c}a^2\\ a^2\ea\right)= {a^2K^2g_1\over 8Ag_2^2}\left(\ba{c}({K\over A-2g_1})^2-1 \\ 1-({K\over A+2g_1})^2\ea\right).
\eeq
Thus we get the solution
\bea
\phi_1&=&{K\over A+2g_1}\sqrt{ {a^2K^2g_1\over 8Ag_2^2} \left(({K\over A-2g_1})^2-1\right) } \\
n_1&=&{K\over A-2g_1}\sqrt{{a^2K^2g_1\over 8Ag_2^2}\left(1-({K\over A+2g_1})^2\right)}\\
\phi_2&=&\sqrt{{a^2K^2g_1\over 8Ag_2^2} \left(({K\over A-2g_1})^2-1\right)} \\
n_2&=&\sqrt{{a^2K^2g_1\over 8Ag_2^2}\left(1-({K\over A+2g_1})^2\right)} .
\eea
The solution is again not particularly illuminating.  We can indeed check that $\phi_i^2+n_i^2=a^2$ for example.  This solution represents the minimum of the potential.  It is easy to see that the order parameter is largely in the $n_1$ direction on the anti-ferromagnetic side while it is largely in the $\phi_2$ direction on the superconductor side.  To find the frequencies of the oscillations about this minimum, we need the matrix of second partial derivatives of the potential.  This is given by
\beq
V^{\prime\prime}=\left(\ba{cccc}A+2g_1+8\lambda\phi_1^2&8\lambda\phi_1n_1&-K&0\\ 8\lambda\phi_1n_1&A-2g_1+8\lambda n_1^2&0&-K\\ -K&0&B-2g_2+8\lambda\phi_2^2&8\lambda\phi_2n_2\\ 0&-K&8\lambda\phi_2n_2&B+2g_2+8\lambda n_2^2\ea\right)
\eeq
where all the fields are evaluated at their values at the minimum.  The frequencies are obtained by diagonalizing this matrix.  This is actually not that difficult, it can be done analytically or by computer.  On the other hand this diagonalization is not very illuminating.  It is possible to work in perturbation theory.  We split the matrix $V^{\prime\prime}=V^{\prime\prime}_0+V^{\prime\prime}_1$ into a zero order part and a perturbation.  The zeroth order part is given by the parts which scale like $\lambda$
\beq
V^{\prime\prime}_0=\left(\ba{cccc}8\lambda\phi_1^2&8\lambda\phi_1n_1&0&0\\ 8\lambda\phi_1n_1&8\lambda n_1^2&0&0\\ &0&8\lambda\phi_2^2&8\lambda\phi_2n_2\\ 0&&8\lambda\phi_2n_2&8\lambda n_2^2\ea\right)
\eeq
while the perturbation is given by the terms of $o (1)$
\beq
V^{\prime\prime}_1=\left(\ba{cccc}A+2g_1&0&-K&0\\ 0&A-2g_1&0&-K\\ -K&0&B-2g_2&0\\ 0&-K&0&B+2g_2\ea\right) .
\eeq
Now finding the eigenvalues and eigenvectors of $V^{\prime\prime}_0$ is trivial.  There are two heavy, massive modes and two Goldstone modes, indeed the Goldstone modes that would be there if all symmetry breaking terms were absent.  It is these Goldstone modes that become the pseudo-Goldstone bosons which interest us.  The perturbation then acts on the degenerate subspace of the two Goldstone modes, lifting the degeneracy and giving both modes a small mass.  They become the pseudo-Goldstone bosons that we are looking for.  Degenerate perturbation theory corresponds simply to diagonalizing the perturbation in the subspace of the degenerate modes. These normalized zero modes are given by
\beq
v_1=(1/a)\left(\ba{c}n_1\\ -\phi_1\\ 0\\ 0\ea\right)\quad v_2=(1/a)\left(\ba{c} 0\\ 0\\ n_2\\ -\phi_2\ea\right) .
\eeq
Then we must compute the matrix elements of the perturbation in the subspace spanned by $v_1$ and $v_2$, this is given by
\bea
V^{\prime\prime}_\Delta &=&\left(\ba{cc}v_1^T V^{\prime\prime}_1v_1&v_1^T V^{\prime\prime}_1v_2\\ v_2^T V^{\prime\prime}_1 v_1&v_2^T V^{\prime\prime}_1v_2\ea\right)\\ &=&(1/a^2)\left(\ba{cc}n_1^2(A+2g_1)+\phi_1^2(A-2g_1)&-K(n_1n_2+\phi_1\phi_2)\\ -K(n_1n_2+\phi_1\phi_2)&n_2^2(B-2g_2)+\phi_2^2(B+2g_2)\ea\right) 
\eea
This matrix is also trivial to diagonalize.  Once again, it is not particularly illuminating to see the eigenspectrum.  It is clear that the mixing between the modes is controlled by $K$.  If we take $K\rightarrow 0$, there is no mixing.  The spectrum is given by
\beq
\omega^2=\alpha\pm\sqrt{\beta^2+\gamma^2}
\eeq
where
\bea
\alpha&=&(1/2a^2)\left( n_1^2(A+2g_1)+\phi_1^2(A-2g_1)+n_2^2(B-2g_2)+\phi_2^2(B+2g_2)\right)\\
\beta&=&(1/2a^2)\left( n_1^2(A+2g_1)+\phi_1^2(A-2g_1)-n_2^2(B-2g_2)-\phi_2^2(B+2g_2)\right)\\
\gamma&=&-(1/2a^2)K(n_1n_2+\phi_1\phi_2) .
\eea
The excitation of these pseudo-Goldstone modes correspond to tunneling across the junction of the electron fluid.  On the superconducting side, it might comprise of standard Cooper pairs, however the mechanism of high temperature superconductivity is not presently understood.  On the anti-ferromagnetic side it is an spin ordered state of the electron fluid which is being excited.  It would be interesting to observe the currents corresponding to this exchange.

\section{Conclusions}
We have shown how to formulate the Josephson effect in the field theoretic language of effective Lagrangians.   This allowed for a straightforward generalization to the theatre of non-abelian symmetries and the corresponding non-abelian Josephson effect.   We find that the Josephson effect corresponds to excitations of pseudo-Goldstone bosons.  The field theoretic description of a superconductor requires that the symmetry is spontaneously broken, which gives rise to Goldstone bosons.  For two uncoupled superconductors, or generalized systems which could undergo the Josephson effect, there is a doubling of the symmetries, simply because all the symmetries of the model occur separately in each distinct superconductor.   Since there is a doubling of the symmetry, there is a consequent doubling of the number of Goldstone bosons produced.  Coupling the two superconductors together explicitly breaks the doubled symmetry to its diagonal subgroup.  This coupling is assumed to be weak, as is the case in the usual Josephson effect.  Hence the would be Goldstone bosons of the softly broken symmetry become slightly massive, and are called pseudo-Goldstone bosons.  

We have demonstrated our formalism for the usual abelian Josephson effect and for two generalizations to non-abelian symmetries.  The first is $U(2)\times SO(3)$ and the second application is to the $SO(5)$ model for the complex of the high temperature superconductivity and anti-ferromagnetism.    

\section{ Acknowledgments }
We thank NSERC of Canada for financial support.


\end{document}